\documentclass{article}
\usepackage{hyperref}
\usepackage{geometry}
\usepackage{footnote}
\usepackage{url}
\title{Future and AI-Ready Data Strategies}
\author{Hamidah Oderinwale\footnote{McGill University, hamidah.oderinwale@mail.mcgill.ca}, Shayne Longpre\footnote{MIT Media Lab, slongpre@media.mit.edu}}
\date{July 17, 2024}
\begin{document}
\maketitle
\section*{Introduction}
The following is a response to the U.S. Department of Commerce's Request for Information (RFI) regarding AI and Open Government Data Assets. First, we commend the Department for its initiative in seeking public insights on the organization and sharing of data. To facilitate scientific discovery and advance AI development, it is crucial for all data producers, including the Department of Commerce and other governmental entities, to prioritize the quality of their data corpora. Ensuring data is accessible, scalable, and secure is essential for harnessing its full potential.
In our response, we outline best practices and key considerations for AI and the Department of Commerce's Open Government Data Assets. To define the latter, we reference the formal legal definition: 
\begin{quote}
    the term ``open Government data asset'' means a public data asset that is— (A) machine-readable; (B) available (or could be made available) in an open format; (C) not encumbered by restrictions, other than intellectual property rights, including under titles 17 and 35, that would impede the use or reuse of such asset; and (D) based on an underlying open standard that is maintained by a standards organization.
\end{quote}
\section*{The Ecosystem}
In 2022, the Open Government Data Act was issued. The Act mandates all federal agencies to “publish their information online as open data, using standardized, machine-readable data formats, with their metadata included in the Data.gov catalog. Then, 2023 became the Year of Open Science. Notably, the NIH’s Data Management and Sharing Policy went into effect at the start of the year, and the NSF invested \$12.5 million into their Findable, Accessible, Interoperable, and Reusable Open Science Research Coordination Networks (FAIROS-RCN) program.\footnote{\url{https://crsreports.congress.gov/product/pdf/IF/IF12299}}\\\\
The government is taking note of how to best manage and use data, especially as AI development needs grow. Epoch AI—a research institute focused on AI—forecasts that LLMs will expend all publicly available data by 2032, at the latest.\footnote{\url{https://epochai.org/blog/will-we-run-out-of-data-limits-of-llm-scaling-based-on-human-generated-data}} If models are overtrained, then there could be no data “left” by 2026. As data becomes a bottleneck for model development, new datasets will become increasingly valuable. The DOC has already demonstrated its commitment to producing more open government data assets. However, as the government takes note, we see a similar shift in priorities within the academic and industry ML research community, especially as AI development needs grow. To encourage the development and discussion of benchmarks and datasets, NeurIPS—one of the most recognized academic conferences in the field—introduced its Datasets and Benchmarks track in 2021.\footnote{\url{https://neurips.cc/Conferences/2022/CallForDatasetsBenchmarks}} With this, we recognize the opportunity at hand to help the DOC build future-ready, AI-ready data infrastructure.
\section*{Challenges and Paths Forward}
One example of what is possible at scale is of a policy analyst working with a tool built on the National Labor Relations Act database. It lets data consumers automatically track different variables. In this case, the database was set to automatically track recent Supreme Court decisions. Then, the UI created an AI-generated summary of key rulings.\footnote{\url{https://GitHub.com/labordata/nlrb-data/tree/main?tab=readme-ov-file}} These capabilities demonstrate how leveraging AI with open government data assets can streamline access to and verification of important information directly from the source.\\\\
Metadata is essentially data about data, providing information such as the origin, context, and structure of data assets. Metadata standards are more useful when they are consistent across systems or, in this case, datasets. To improve its metadata standards, the DOC should consider adopting temporal metadata tags: saving versions and marking when a dataset was created, edited, or deleted. This is helpful for research reproducibility, like a social scientist completing a longitudinal study with a live dataset from a DOC agency. It allows researchers to reproduce analyses based on specific data versions, compare different iterations of a dataset, and understand how annotations or corrections have evolved. Temporal metadata would be useful to adopt alongside dataset publishing platforms like GitHub, which provides an interface to interact with different versions.\footnote{\url{https://x.com/MattBruenig/status/1806842258638901505}} Temporal metadata is helpful for formal version releases of a dataset and building DOC-specific tools.
\subsection*{Annotation}
The DOC should invest in building AI-ready datasets. A feature of a high-quality, AI-ready dataset is labels. Labels or "detailed metadata" help automatic AI or scientific analysis programs differentiate relevant from irrelevant records, potential data biases, omissions, or errors.\\\\
Furthermore, labels help models learn by giving them guidelines for what is correct. A model can then train into the most accurate model faster than it would otherwise. For the unlabeled data that the DOC agencies (such as NOAA, NIST, and others) might generate or data with low-information labels and insufficient metadata, they should consider hiring workers and employing annotation services to label and annotate it properly.
\subsection*{Sharing data on public repositories on the web}
Publishing DOC datasets on public repositories would make them easily searchable and discoverable for developers. Platforms like Hugging Face, GitHub, Google, and Meta Research's Papers with Code are engineered for searchability, navigation, and discoverability, where developers can find the datasets most useful for their use cases.
The most notable platform is GitHub. GitHub has logged collaborative version control so users can share their code and data repositories and see logged changes. The community can also contribute to repositories, while the owner retains control over which contributions are accepted.\footnote{\url{https://huggingface.co/docs/hub/en/datasets-adding}} Hugging Face, the largest ML dataset host and built-in search engine, would help DOC datasets be found on its platform and in public search engine results. Additionally, making DOC assets available in public data repositories would help them be seen by those who need them. One is the Hugging Face data repository, which has metadata UI and dataset cards. Their metadata support lets developers add information about their dataset, including its licensing.\footnote{\url{https://paperswithcode.com/datasets}} Papers with Code is a similar platform and another highly-frequented webpage for datasets.\\\\
An example is Google Cloud’s Public Datasets. Their repository includes government and DOC datasets, including Census Bureau US Boundaries datasets and the Google Patents Research Database.\footnote{\url{https://console.cloud.google.com/marketplace/product/google_patents_public_datasets/uspto-oce-claims}} While there are a number of USPTO datasets, they have not been updated. For example, the USPTO OCE Patent Claims Research dataset only shares the U.S. patents granted between 1976 and 2014 and U.S. patent applications published between 2001 and 2014.
Additionally, while other repositories are not pay-walled, Google’s is one of the most comprehensive, with government and public datasets. It is worth considering partnerships to offer them for free through these platforms as well.
\subsection*{Special Tabulations}
Presently, sponsors or requesters can ask for special tabulations—requests for (snapshots of) unpublished but publicly available data from government organizations, the US Census Bureau being one of them. An example of unpublished publicly accessible data is the data approved to be shared through FOIA (Freedom of Information Act) requests–found in mandated FOIA Reading Rooms.\\\\
Special tabulations are fulfilled by statistical agencies housed at federal agencies. These agencies are responsible for providing requested datasets, typically snapshots from a larger pool of data.\\\\
Sponsors or requesters can ask for special tabulations—requests for unpublished yet publicly available data offered by statistical agencies within federal agencies. While the service and data requested are valuable, the current process is mainly manual. Requests are made via email, where the requester provides a summary of their request, the reason for it, the desired output file format, and the ideal delivery date. Each case is handled individually, with emails being read and processed on a per-case basis, which is inefficient.\\\\
The process can be streamlined and made more public and transparent, where requests for datasets could be documented and treated as integral datasets themselves. Instead of emails as a way to submit requests, they could be made through a form connected to a database. Forms would have token fields. Token fields let users enter multiple values, such as keywords or tags, separated by delimiters like commas or spaces. Each entry within the field is treated as a distinct ``token.'' Then, a system could either be developed from scratch or an existing third-party solution could be adopted to make the database searchable. This data can then be synchronized with a cloud provider, like AWS, to manage requests and add additional functionality to the system. Further development could automate the tabulation development process, parse requests, and generate
the dataset.\footnote{\url{https://services.google.com/fh/files/misc/public_datasets_one_pager.pdf}}
\section*{Patent Data and Structural Metadata for Semantic Search}
The DOC is home to valuable data. Patent data from the DOC’s USPTO looks especially promising. For example, patent landscaping—documenting and analyzing patent trends over time– is underexplored relative to how effective it can be at helping policymakers understand and regulate emerging technology over time. Research from Mirac Suzgun et al.\footnote{\url{https://www.sciencedirect.com/science/article/abs/pii/S0172219014001367}} shows structural metadata, specifically “inventor-submitted versions of patent applications” instead of the final versions of granted patents. This metadata helped the researchers learn more from the same dataset and leverage NLP techniques to speed up and deepen their research analysis.\\\\
Semantic search can improve the discovery of government data assets in existing repositories and assist the DOC in creating new ones. Often, keywords must be exact to find a relevant dataset, making it difficult for people to find what they need. Semantic search relies on structural metadata to understand the context and meaning of search queries, which can be more conversational and aligned with the user's intent.
\\\\While traditional search engines focus on finding exact keyword matches within documents, semantic search uses structural metadata and advanced algorithms to comprehend the meaning and relationships between concepts, providing more relevant and accurate results.\footnote{\url{https://www.csis.org/analysis/what-can-patent-data-reveal-about-us-china-technology-competition}}
\subsection*{Benchmarking Generative AI}
Chatbots could be very helpful to government workers, helping them with the tasks they have to accomplish day-to-day: parsing and synthesizing dense texts, trying to make sense of their readings and communicating their findings and ideas to stakeholders. To safely take advantage of the capabilities of Generative AI, such as creating AI-based summaries, the output should be reliable. For this task in particular, a summary of a piece of legal text to be cited should be accurate and free of hallucinations.\\\\
One way to do this is through benchmarks which are tests for model capabilities. Some examples include MMLU (Massive Multitask Language Understanding),\footnote{\url{https://paperswithcode.com/sota/multi-task-language-understanding-on-mmlu}} ChatBot Arena\footnote{\url{https://huggingface.co/spaces/lmsys/chatbot-arena-leaderboard}}, and GLUE (General Language Understanding Evaluation).\footnote{\url{https://gluebenchmark.com/leaderboard/}}\\\\
As it is, we are not aware that there are no benchmarks that test for policy-specific knowledge, especially DOC-related knowledge. There is an opportunity to pursue this. However, it will be important to address some ways in which the most widely used benchmarks fall short: developing fine-tuned models for policy that are as robust as possible. Recent work has found that data contamination, where benchmarks will leech into the models they are trying to test, is common and jeopardizes our understanding of SOTA capabilities. As an example of what could be created, Legalbench is a legal benchmark that tests the legal reasoning abilities of LLMs.\footnote{\url{https://hazyresearch.stanford.edu/legalbench/}}
\section*{Data Formats} How federal agencies and the DOC share the format of their data is instrumental to its accessibility and therefore how much it’s used. Instead of providing PDFs, which make it difficult to accurately extract the underlying text, images, or other content, agencies should share raw data that’s not been altered and make sure that processed data is in a machine-readable format. For example, a Bureau of Economic Analysis’ (BEA) GDP data collection was a mix of PDFs and XSLX files, which means that there was no machine-readable representation of their text-based data (i.e., technical reports). While HTML—the markup language that sites are written in—can technically be read by machines, it isn’t easily processed by them. Therefore, written content should also be available in data formats like JSON and XML that are designed for computer processing.\footnote{\url{https://www.bea.gov/news/2024/gross-domestic-product-third-estimate-corporate-profits-revised-estimate-and-gdp-industry}}
\section*{Publishing DOC-Authored Documents}
Online document editors and PDFs are often static and should reflect changes better than they currently do. They should have versioning control and persistent identifiers, making it difficult to keep track of updates and ensure accurate references. Publishing DOC-authored documents on a platform like PubPub—an open-source community publishing platform by the Knowledge Futures Group—offers several advantages. PubPub supports the assignment of DOIs, which is not possible with online document editors. Tools like PubPub, with versioning and CrossRef DOIs, keep links persistent and updated even with changes to the original content.\footnote{\url{https://www.pubpub.org/}}
DOIs ensure that each document has a unique and permanent identifier, making it easier to reference and cite accurately. Additionally, PubPub provides machine-readable formats, making the content easily accessible for AI and data processing. The platform allows seamless sharing through links instead of files, facilitating broader distribution and collaboration.\\\\
Furthermore, PubPub\footnote{\url{https://www.pubpub.org}} supports dynamic versioning, allowing for transparent updates and revisions over time, which is beneficial for maintaining document accuracy and relevance. This feature lets users see a document's evolution, track changes, and access previous versions if needed. PubPub can also be a sandbox, providing a flexible environment for drafting, experimenting, and refining documents before final publication.
\section*{Relevant Work}
Now, we outline the initiatives and resources relevant to data-centric machine learning. Firstly, the Data-centric AI Resource Hub that came out of the 2021 NeurIPS Data-Centric AI (DCAI) workshop.\footnote{\url{https://datacentricai.org/}} The Data Provenance Initiative has worked on auditing datasets used to train machine learning models. In their first audit in 2023, they reviewed over 1,800 text datasets, collecting essential details such as the origin, creators, and licensing conditions of each dataset. This effort ensures that the data used in machine learning is well-documented and responsibly handled.\footnote{\url{https://www.dataprovenance.org/}}\\\\
Datasheets for datasets, introduced in 2018, are now an established best practice for ML dataset development.\footnote{\url{https://arxiv.org/pdf/1803.09010}} With datasheets, a dataset creator answers a series of questions in a text file that describes the motivations, process, intentions for distribution, impact, and disclosure of any issues with their dataset.\\\\
The Foundation Model Development Cheatsheet is ``a succinct guide, prepared by foundation model developers for foundation model developers.'' Data-relevant topics include catalogues of data sources, tools for data preparation, and resources for Data Search, Analysis, and exploration.\footnote{\url{https://fmcheatsheet.org/}} The MLCommons’ Croissant Initiative also developed an ML metadata schema built on schema.org.\footnote{\url{http://schema.org}} Lastly, work from the DMLR workshop at 2023 International Conference on Machine Learning (ICML) gives a comprehensive overview of many of the initiatives and key research projects in the space.\footnote{\url{https://openreview.net/pdf?id=2kpu78QdeE}}
\end{document}